\documentclass[prb,superscriptaddress,amsfonts,twocolumn,showpacs,floatfix]{revtex4}

\usepackage{amsmath}
\usepackage{amsmath}
\usepackage{amsfonts}
\usepackage{amssymb}
\usepackage{txfonts}
\usepackage{bm}
\usepackage{tabularx}
\usepackage{graphicx,color}

\def\Vec#1{\bm{#1}}
\def\Hc2{H_\mathrm{c2}}
\def\Tc{T_\mathrm{c}}

\begin{document}

\title{Superconducting gap structure of CeIrIn$_5$ from field-angle-resolved measurements \\of its specific heat}

\author{Shunichiro Kittaka}
\affiliation{Institute for Solid State Physics, University of Tokyo, Kashiwa, Chiba 277-8581, Japan}
\author{Yuya Aoki}
\affiliation{Institute for Solid State Physics, University of Tokyo, Kashiwa, Chiba 277-8581, Japan}
\author{Toshiro Sakakibara}
\affiliation{Institute for Solid State Physics, University of Tokyo, Kashiwa, Chiba 277-8581, Japan}
\author{Akito Sakai}
\affiliation{Institute for Solid State Physics, University of Tokyo, Kashiwa, Chiba 277-8581, Japan}
\author{Satoru Nakatsuji}
\affiliation{Institute for Solid State Physics, University of Tokyo, Kashiwa, Chiba 277-8581, Japan}
\author{Yasumasa Tsutsumi}
\affiliation{Department of Physics, Okayama University, Okayama 700-8530, Japan}
\author{Masanori Ichioka}
\affiliation{Department of Physics, Okayama University, Okayama 700-8530, Japan}
\author{Kazushige Machida}
\affiliation{Department of Physics, Okayama University, Okayama 700-8530, Japan}

\date{\today}

\begin{abstract}
In order to identify the gap structure of CeIrIn$_5$,
we  measured field-angle-resolved specific heat $C(\phi)$ by conically rotating the magnetic field $H$ around the $c$ axis at low temperatures down to 80~mK.
We revealed that $C(\phi)$ exhibits a fourfold angular oscillation, 
whose amplitude decreases monotonically by tilting $H$ out of the $ab$ plane.
Detailed microscopic calculations based on the quasiclassical Eilenberger equation confirm that
the observed features are uniquely explained by assuming the $d_{x^2-y^2}$-wave gap.
These results strongly indicate that CeIrIn$_5$ is a $d_{x^2-y^2}$-wave superconductor and 
suggest the universal pairing mechanism in Ce$M$In$_5$ ($M=$ Co, Rh, and Ir).
\end{abstract}


\maketitle

The heavy-fermion systems Ce$M$In$_5$ ($M=$ Co, Rh, and Ir) have been extensively studied 
because they exhibit unconventional superconductivity near the antiferromagnetic (AF) quantum critical point (QCP).
Especially, much effort has been expended to identify the superconducting (SC) gap structure, a challenging issue that is closely related to the identification of the pairing mechanism.
Recently, field-angle-resolved experiments along with theoretical works have confirmed that
CeCoIn$_5$ is a $d_{x^2-y^2}$-wave superconductor.\cite{An2010PRL,Izawa2001PRL,Vorontsov2007PRB}
Because of the $d_{x^2-y^2}$-wave gap and the proximity of the system to the AF state,
it is now widely accepted that the pairing interaction in CeCoIn$_5$ is AF spin fluctuations.
Pressure induced superconductivity in CeRhIn$_5$ has also been examined by field-angle-resolved specific heat measurements and it is argued to have the same pairing symmetry.\cite{Park2008PRL}

By contrast, the possibility of a different pairing mechanism has been suggested for CeIrIn$_5$ 
due to its unusual behavior of the SC phase.\cite{Holmes2007JPSJ}
The transition temperature $\Tc$, which is 0.4~K at $P$=0,  increases to 0.8~K under a high pressure of $P$=2.1~GPa,\cite{Kawasaki2005PRL}
although strong AF fluctuations existing at ambient pressure are rapidly suppressed by increasing $P$ as the system is further pushed away from a hypothetical AF QCP. 
Moreover, when Rh  is doped to Ir sites,  $\Tc$ shows a cusp-like minimum before reaching the maximum $\Tc$ of $\sim$1~K around the onset  of an AF state.\cite{Nicklas2004PRB,Kawasaki2006PRL}
These observations suggest the presence of two distinct SC phases in CeRh$_{1-x}$Ir$_x$In$_5$.

While several attempts have been made to uncover the gap structure of CeIrIn$_5$, two conflicting possibilities have remained.
Kasahara $et$ $al$.~\cite{Kasahara2008PRL} reported a fourfold angular oscillation in the thermal conductivity $\kappa(\phi)$ when $H$ is rotated in the basal plane, 
and attributed its origin to the vertical line nodes of the $d_{x^2-y^2}$-wave gap.
On the other hand, Shakeripour $et$ $al$.~\cite{Shakeripour2007PRL,Shakeripour2010PRL} examined the effect of impurity scattering of $\kappa(T)$ for a current parallel and perpendicular to the $c$ axis, and 
proposed  the gap function of CeIrIn$_5$ to be either $k_z$ or $k_z(k_x+ik_y)$, both of which have a horizontal line node only on the equator, in sharp contrast to the $d_{x^2-y^2}$-wave gap.
At present, neither of these two possibilities can be ruled out.
The temperature variation of the anisotropy $\kappa_c/\kappa_a$ in Ref.~\onlinecite{Shakeripour2007PRL} can be explained within the  $d_{x^2-y^2}$ symmetry 
if the Fermi surface has small deviations from the cylindrical symmetry,\cite{Vekhter2007PRB} 
whereas the fourfold oscillation in $\kappa(\phi)$ can be explained by the horizontal line node gap 
if an in-plane anisotropy of the effective mass and/or minima exist in the azimuthal variation of the gap amplitude.\cite{Shakeripour2010PRL} 
Thus, the gap structure of CeIrIn$_5$ has remained  controversial.

In order to settle the controversy over the gap structure of CeIrIn$_5$, we have performed field-angle-resolved specific heat $C(\phi, \theta)$ measurements down to 80 mK. 
The $C(\phi)$ measurements, where $\phi$ denotes the in-plane azimuthal angle of $H$, have proven to be quite useful to determine the direction of nodes in the momentum space of bulk superconductors.\cite{Sakakibara2007JPSJ}
Here we extend the method to measure the polar angle $\theta$ dependence of $C(\phi)$ by which a detection of the horizontal line node can be made.
We revealed that $C(\phi, \theta)$ exhibits a clear fourfold oscillation as a function of $\phi$ with $H$ rotated around the $c$ axis,  and its amplitude is monotonically suppressed by tilting $H$ out of the $ab$ plane (decreasing $\theta$ from 90$^\circ$).
The results are compared with the microscopic theory by solving the quasiclassical Eilenberger equation self-consistently, and are found to be in good agreement with the $d_{x^2-y^2}$-wave gap, 
but are in a sharp contrast with the behavior predicted for the horizontal line node gap.

The single crystal of CeIrIn$_5$ used in the present study ($\Tc=0.4$ K, 40.4 mg) was grown by the self-flux method.
The specific heat was measured by the relaxation and the standard adiabatic heat-pulse methods in a dilution refrigerator (Oxford Kelvinox AST Minisorb).
Magnetic fields were applied in the $xz$ plane by using a vector magnet consisting of 
horizontal split-pair (5~T) and vertical solenoid (3~T) coils.
By rotating the refrigerator around the $z$ axis using a stepper motor mounted at the top of a magnet Dewar,
three dimensional control of the magnetic field direction is achieved.
We confirmed that the addenda contribution was always less than 5\% of the sample specific heat and had no field-angle dependence.
An accurate ($< 0.1^\circ$) and precise ($< 0.01^\circ$) field alignment with respect to the crystalline $ab$ plane was accomplished 
by making use of the $C(\theta)$ data which reflects the tetragonal anisotropy of $H_\mathrm{c2}$.


The specific heat of CeIrIn$_5$ at zero field is known to exhibit an upturn on cooling below about 0.1 K due to a quadrupole splitting of the $^{115}$In ($I=9/2$) and $^{191,193}$Ir ($I=3/2$) nuclear spins.\cite{Petrovic2001EPL,Movshovich2001PRL}
In this paper, the nuclear Schottky contribution was subtracted from the data assuming $C_\mathrm{n}=(a_0+a_1H^2)/T^2$,
where $a_0$ was adjusted so that the resulting electronic contribution $C_\mathrm{e}=(C-C_\mathrm{n})$ at $H=0$ becomes proportional to $T^2$ at low $T$.
The value of $a_0$ ($=0.38$ mJ K/mol) thus obtained was in good agreement with the calculated one ($\sim 0.34$ mJ K/mol) 
using the parameters determined from the $^{115}$In nuclear quadrupole resonance experiment.\cite{Zheng2001PRL,Kohori2001PRB}
The coefficient $a_1$ was calculated from the nuclear Zeeman splitting.
It can be shown that  $C_\mathrm{n}$ is independent of the in-plane field orientation.\cite{An2010PRL}
Because of the nuclear Schottky contribution, the  $C(\phi,\theta)$ measurements were limited to the temperature range $T\geq$~80~mK.

\begin{figure}
\begin{center}
\includegraphics[width=3.2in]{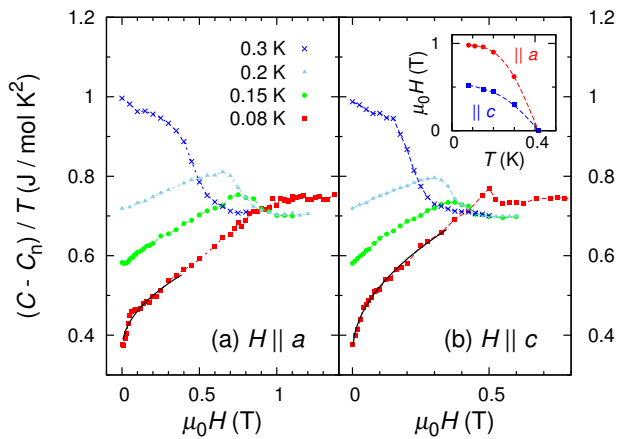}
\end{center}
\caption{
(Color online) 
Magnetic field dependence of the nuclear subtracted specific heat divided by temperature $C_\mathrm{e}/T$ 
for (a) $H \parallel [100]$ and (b) $H \parallel [001]$.
The solid lines represent fits to the data by $a\sqrt{H}+b$.
The inset in (b) shows the temperature dependence of the upper critical field determined from the present study.
}
\label{fig1}
\end{figure}

Figures 1(a) and 1(b) show $C_\mathrm{e}(H)$ for $H \parallel a$ and $H \parallel c$, respectively.
As represented by the solid lines, $C_\mathrm{e}(H)$ is proportional to the square root of $H$ at low $H$ and low $T$.
This behavior supports the presence of line nodes in the SC gap.
The $C_\mathrm{e}(H)$ behavior near $\Hc2$ is in good agreement with the calculated result for a two-dimensional $d$-wave superconductor 
with a relatively small Pauli paramagnetic parameter $\mu=0.86$.\cite{Ichioka2007PRB}
At 80~mK for $H \parallel a$, a cusp-like structure is observed at 0.06 T,
which might originate from the multi-gap superconductivity, 
as reported in Sr$_2$RuO$_4$ (Ref. \onlinecite{Deguchi2004JPSJ}) and MgB$_2$.\cite{Boaknin2003PRL}

Based on the $C_\mathrm{e}(H)$ data, we determined the upper critical field $H_\mathrm{c2}(T)$, as plotted in the inset of Fig. 1(b).
At the lowest temperature 80~mK, $H_\mathrm{c2}$ of the present sample is 1.0 T for $H \parallel a$ and 0.55 T for $H \parallel c$.
The reduced ratio $\alpha=-H_\mathrm{c2}/(T_\mathrm{c}\mathrm{d}H_\mathrm{c2}/\mathrm{d}T|_{T=T_\mathrm{c}})$
is estimated to be 0.47 for $H \parallel a$ and 0.54 for $H \parallel c$, 
which agree well with the results of Ref.~\onlinecite{Movshovich2002PhysicaB}.
These results also suggest that the Pauli paramagnetic effect in CeIrIn$_5$ is relatively weak compared with that in CeCoIn$_5$ [$\alpha=0.26$ (Ref. \onlinecite{Movshovich2002PhysicaB})].

Figure 2 shows the $\phi$ dependence of $C_\mathrm{e}/T$ obtained by rotating $H$ in the $ab$ plane,
where $\phi$ is measured from the [100] direction.
For each data point, the $C_\mathrm{e}(\phi)$ value was determined by an average of ten successive measurements.
The error bar is estimated to be several mJ/mol$\cdot$K$^2$ for $T \ge 110$~mK, 
while it becomes about 10~mJ/mol$\cdot$K$^2$ at 80~mK due to the large nuclear contribution.
A clear fourfold oscillation was observed in $C_\mathrm{e}(\phi)$ in the wide $T$ and $H$ region.
We carefully confirmed the absence of a twofold oscillation in $C_\mathrm{e}(\phi)$ 
which guarantees the accurate and precise alignment of the magnetic field with respect to the $ab$ plane.
In this intermediate $T$ regime ($0.2 \le T/\Tc \le 0.5$), 
$C_\mathrm{e}(\phi)$ becomes minimum in fields along the $\langle 100 \rangle$ directions.

\begin{figure}
\begin{center}
\includegraphics[width=3.2in]{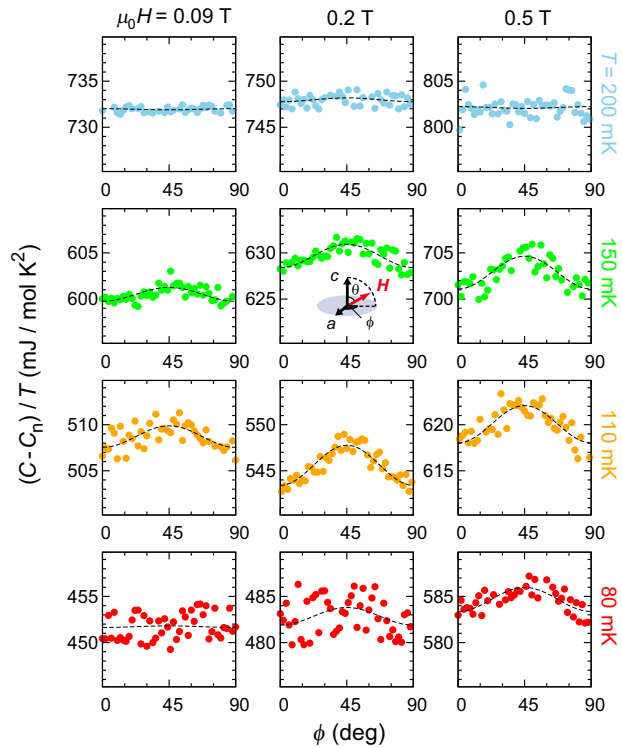}
\end{center}
\caption{
(Color online) 
Variations of $C_\mathrm{e}/T$ as a function of the azimuthal angle $\phi$ between the [100] axis and the magnetic field applied in the $ab$ plane ($\theta=90^\circ$).
The dashed lines represent fits to the data by $C_0(T)+C_H(T,H)(1-A_4\cos4\phi)$.
}
\label{fig2}
\end{figure}

To characterize the $H$ and $T$ variations of the fourfold oscillation, we fit the data to the expression 
\begin{equation}
C_\mathrm{e}(T,H,\phi)=C_0(T)+C_H(T,H)(1-A_4\cos4\phi)
\label{eq:Cphi}
\end{equation}
as represented in Fig.~2 by the dashed lines.
Here, $C_0$ and $C_H$ are the zero-field and field-dependent components of $C_\mathrm{e}$, respectively, and
$A_4$ is the amplitude of the fourfold oscillation normalized by $C_H$. 
Figure 3(a) shows the $T$ dependence of $A_4$ at $\mu_0H$=0.09, 0.2, and 0.5~T.
We found that $A_4(T,H)$ exhibits a peak at around $0.3T_\mathrm{c}$ and $0.15\Hc2$, and rapidly decreases down to zero around $0.2T_\mathrm{c}$. 
This is a feature that, to the best of our knowledge, has not been observed in previous $\kappa(\phi)$ measurements done mainly above $0.4T_\mathrm{c}$,\cite{Kasahara2008PRL} and strongly suggests the existence of fourfold vertical line nodes.
These features can be seen more clearly by the contour plot of $A_4(T,H)$ in Fig. 3(b).

\begin{figure}
\begin{center}
\includegraphics[width=3.08in]{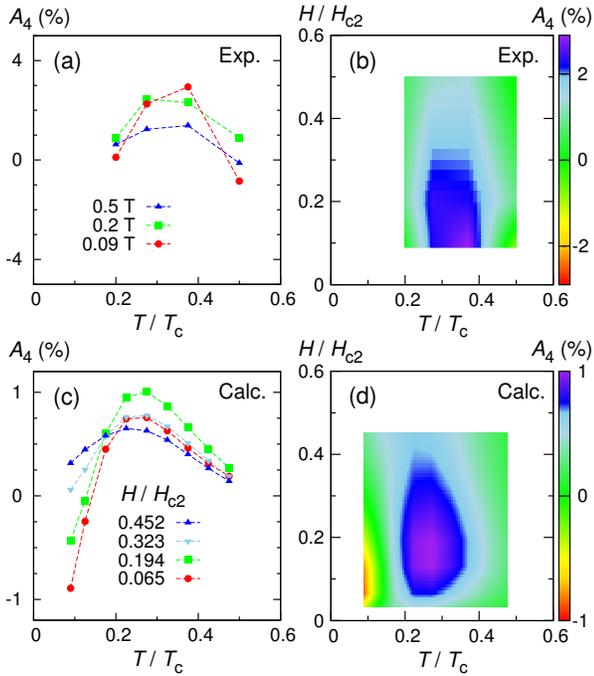}
\end{center}
\caption{
(Color online) 
Temperature dependence of the normalized fourfold amplitude $A_4$ at several fields 
obtained by (a) the present experiment and (c)~the microscopical calculations assuming the $d_{x^2-y^2}$-wave gap.\cite{Hiragi2010JPSJ}
(b), (d) Contour plots of $A_4(T,H)$ using the same data in (a) and (c), respectively.
Here, $T_\mathrm{c}=0.4$ K and $\mu_0\Hc2=1$ T.
}
\label{fig3}
\end{figure}

\begin{figure}
\begin{center}
\includegraphics[width=3.2in]{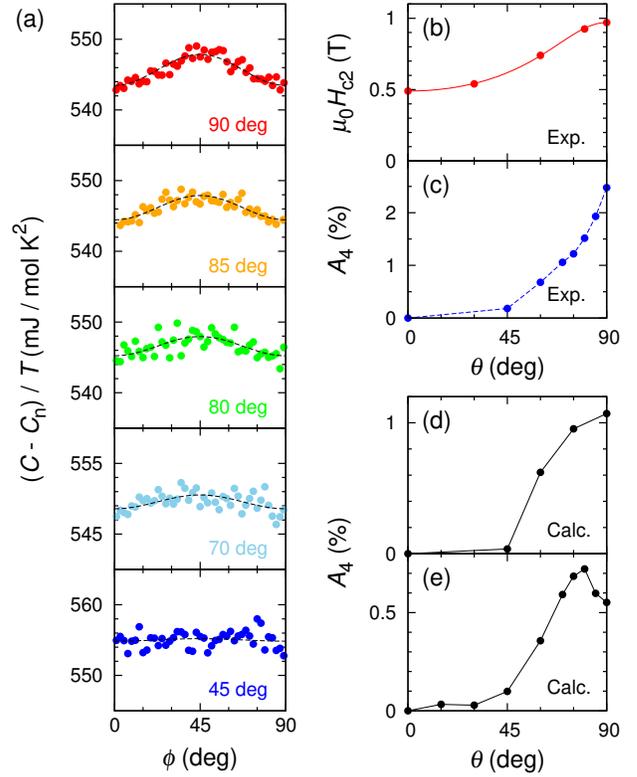}
\end{center}
\caption{
(Color online) 
(a) Field-angle $\phi$ dependence of $C_\mathrm{e}/T$ in a conically rotating magnetic field with a strength of $0.2H_\mathrm{c2}(\theta)$ 
at several fixed polar angles $\theta$, measured at 110~mK.
The dashed lines are fits to the data by the expression $C_0(T)+C_H(T,H)(1-A_4\cos4\phi)$.
(b) and (c) show the polar angle $\theta$ dependence of $H_\mathrm{c2}$ and $A_4$ at 110 mK, respectively.
The calculated $\theta$ dependence of the zero-energy DOS anisotropy 
$\left[N_0(H\parallel {\rm antinode})-N_0(H\parallel {\rm node})\right]/2N_0(H)$
is shown in (d) and (e) for the order parameters $k_x^2-k_y^2$ and $k_z(k_x^2-k_y^2)$, respectively.
}
\label{fig4}
\end{figure}

The $\phi$ rotation experiment alone, however, cannot rule out the possibility of the horizontal line node as claimed in Ref.~\onlinecite{Shakeripour2010PRL}.
In order to solve this issue, 
we investigated $C_\mathrm{e}(\phi)$ by conically rotating $H$ at several fixed $\theta$,
where $\theta$ denotes the polar angle between $H$ and the $c$ axis.
The results obtained at 110~mK are shown in Fig.~4(a), where the dashed lines are the fit to Eq.~(\ref{eq:Cphi}).
In these measurements, the intensity of $H$ is adjusted for each $\theta$ to satisfy $H/H_\mathrm{c2}(\theta)=0.2$ 
in order to avoid  unfavorable effects of the tetragonal anisotropy in $H_\mathrm{c2}$.
The $\theta$ dependence of $H_\mathrm{c2}$ was experimentally determined from the $C_\mathrm{e}(H)$ measurements at $T$=110~mK and is  shown in  Fig.~4(b) by the solid circles.
The solid line in the figure is a fit to the 
Ginzburg-Landau anisotropic-effective-mass formula for three-dimensional superconductors \cite{Morris1972PRB}
$\Hc2(\theta)=H_{\mathrm{c2} \parallel c}/(\cos^2\theta+(H_{\mathrm{c2} \parallel a}/H_{\mathrm{c2} \parallel c})^2\sin^2\theta)^{1/2}$, which reproduces the observed $\theta$ variation of $H_\mathrm{c2}$ satisfactorily.
Figure 4(c) shows the $\theta$ dependence of $A_4$ obtained from the results in Fig.~4(a).
We found that $A_4$ decreases gradually and monotonically
with decreasing $\theta$ from 90$^\circ$ ($H\perp c$) to 0$^\circ$.

Let us compare the experimental results with microscopic calculations and discuss the SC gap structure of CeIrIn$_5$.
We assume a two-dimensional cylindrical Fermi surface for the main Fermi surface of CeIrIn$_5$ with significant $f$-electron contribution.\cite{Haga2001PRB}
The local density of states (LDOS) of the quasiparticles (QPs) under the rotating $H$ was calculated by solving the quasi-classical Eilenberger equation self consistently. 
Here, we assume the gap function to be $k_x^2-k_y^2$, and no Pauli paramagnetic effect ($\mu=0$) is considered.
The $H$ and $T$ variations of the $A_4$ coefficient were then evaluated from the LDOS~\cite{Hiragi2010JPSJ} and the results are shown in Figs.~3(c) and 3(d) (contour plot).
In the low $H$ ($<0.2H_{\rm c2}$) and low $T$ ($<0.12T_{\rm c}$) region, $A_4$ becomes negative implying that the $C_\mathrm{e}(\phi)$ oscillation has minima along the nodal directions ($\langle 110 \rangle$).
In this regime, $A_4$ is approximately proportional to the anisotropy of the zero-energy DOS $N_0$ [$|A_4| \propto 1-N_0(H \parallel {\rm node}) / N_0(H \parallel {\rm antinode})$]~\cite{Miranovic2005JPC} that can be intuitively understood by the Doppler-shift effect of the QPs on the circulating supercurrent around the vortices.
The Doppler shift is given by $\delta = m_\mathrm{e}\Vec{v}_\mathrm{F} \cdot \Vec{v}_\mathrm{s}$, where $m_\mathrm{e}$ is the electron mass, $\Vec{v}_\mathrm{F}$ is the velocity of the QP and $\Vec{v}_s$ is the local superfluid velocity that is perpendicular to $H$.
For a superconductor with line nodes, $N_0$ is enhanced by the Doppler shift of the nodal QPs, giving rise to the $H^{1/2}$ behavior of $C_\mathrm{e}(H)$. 
If $H$ is in a nodal direction, then those QPs cannot contribute to $N_0$ because $\delta$ vanishes. As a consequence, $N_0$ exhibits an angular oscillation with $N_0(H \parallel {\rm node}) <N_0(H \parallel {\rm antinode})$.

As $T$ increases, contributions of the finite-energy DOS to $A_4$ become relevant.\cite{Vorontsov2006PRL,Vorontsov2007PRB}
The detailed calculations tell us that the sign of $A_4$ changes around $0.15T_{\rm  c}$ as shown in Fig.~3(c), and 
$C_\mathrm{e}(\phi)$ takes minima for $H\parallel$ antinodal directions in the intermediate $T$ range.
These theoretical results well reproduce our experimental data as can be seen by comparing the contour plots in Figs.~3(b) and 3(d).
Especially, the peak position and the line of the sign change in $A_4(T,H)$ agree remarkably with each other.
The agreement becomes worse if a relatively strong Pauli effect is introduced ($\mu$=2 -- see Fig. 11 in Ref. \onlinecite{Hiragi2010JPSJ}).
This fact is compatible with the weak Pauli effect ($\mu < 1$) expected from $C_\mathrm{e}(H)$.
Although the low-$T$ Doppler-shift-predominant region could not be reached in our experiment, our data strongly indicate nodes along $\langle 110 \rangle$ directions.

In order to provide further evidence of the $d_{x^2-y^2}$-wave gap in CeIrIn$_5$, we calculated the $\theta$ dependence of $A_4$.
Here we compare two types of gap functions $k_x^2-k_y^2$ and $(k_x^2-k_y^2)k_z$,  and the results are shown in Figs. 4(d) and 4(e).
The latter mimics the case of a horizontal line node gap with an azimuthal angular variation of the gap amplitude 
as claimed in Ref.~\onlinecite{Shakeripour2010PRL}.
For the $d_{x^2-y^2}$-wave gap, 
a gradual and monotonic decrease of $A_4$ on decreasing $\theta$ from 90$^\circ$ is expected [Fig.~4(d)], 
whereas for the $(k_x^2-k_y^2)k_z$ gap $A_4(\theta)$ is predicted to have a dip around $\theta=90^\circ$ [Fig. 4(e)].
The reason why $A_4(\theta)$ shows a local minimum at $\theta=90^{\circ}$ for the horizontal line node gap is because there always exist nodal QPs whose momentum is parallel to the field direction, irrespective of $\phi$.
In this case, the anisotropy in the Doppler shift $\delta$ is strongly reduced, leading to the reduction of $A_4$.
It is obvious that the $k_x^2-k_y^2$ gap rather than the horizontal line node gap better reproduces the experimental data in Fig.~4 (c).
In particular, the absence of the dip of $A_4(\theta)$ at $\theta=90^{\circ}$ in our results strongly indicates that CeIrIn$_5$ does not have a horizontal line node at $k_z=0$.
These observations  lead us to conclude that 
the SC gap in CeIrIn$_5$ is of $d_{x^2-y^2}$ type.

In conclusion, we have performed field-angle-resolved specific-heat measurements on CeIrIn$_5$ with $\Tc=0.4$~K by conically rotating the magnetic field around the $c$ axis.
The specific heat exhibits a clear fourfold angular oscillation and 
its dependences on temperature and field is surprisingly well reproduced by the microscopic calculations 
for a $d_{x^2-y^2}$-wave superconductor. 
One of the most important findings in the present study is the monotonic variation 
of the oscillation amplitude with the polar angle of a conically rotating field.
This feature confirms the absence of a horizontal line node on the equator 
and strongly supports the $d_{x^2-y^2}$-wave gap, as in CeCoIn$_5$ and CeRhIn$_5$.
The establishment of the identical gap symmetry in Ce$M$In$_5$ ($M=$ Co, Rh, and Ir) 
indicates the universality of the pairing mechanism in this family and provides important hints for resolving the mechanism of the unique superconductivity.

This work has been partly supported by  
Grants-in-Aid for Scientific Research on Innovative Areas ``Heavy Electrons'' (20102007, 23102705)
from the Ministry of Education, Culture,
Sports, Science and Technology of Japan.

\textit{Note added in proof}.--- 
Recently, Lu \textit{et al}. also reported $C(\phi)$ of CeIrIn$_5$ under pressure down to 0.3 K
and the same conclusion has been reached. \cite{Lu2011PRL}


\end{document}